\documentclass[aps,pre,twocolumn,showpacs,superscriptaddress]{revtex4}

\usepackage{amsmath,amssymb,graphicx,textcomp}

\usepackage{times}
\usepackage{color}

\newcommand{\kl}{\left (}
\newcommand{\kr}{\right )}
\newcommand{\ekl}{\left [}
\newcommand{\ekr}{\right ]}
\renewcommand{\vec}{\mathbf}
\newcommand\diff{\mathrm{d}}

\begin{document}
\bibliographystyle{apsrev}

\title{Persistent correlation of constrained colloidal motion}
\author{Thomas Franosch${}^*$}
\affiliation{Arnold Sommerfeld Center for Theoretical Physics (ASC) and Center for NanoScience (CeNS), Department of Physics, Ludwig-Maximilians-Universit{\"a}t M{\"u}nchen, Theresienstra{\ss}e 37, D-80333 M{\"u}nchen, Germany}
\altaffiliation{Corresponding author. E-mail:
franosch@lmu.de}

\author{Sylvia Jeney}
\affiliation{Institut de Physique de la Mati\`ere Complexe, Ecole
Polytechnique F\'ed\'erale de Lausanne (EPFL), CH-1015 Lausanne,
Switzerland}

\date{\today}

\begin{abstract}
We have investigated the motion of a single optically trapped colloidal particle  close to a limiting wall at time scales where the inertia of the surrounding fluid plays a significant role.
The velocity autocorrelation function exhibits a complex interplay due to the momentum relaxation of the particle, the vortex diffusion in the fluid, the obstruction of flow close to the interface, and the  harmonic restoring  forces due  to the optical trap. We show that already a weak trapping force has a significant impact on the velocity autocorrelation function
$C(t)=\langle v(t)v(0)\rangle$ at times where the hydrodynamic memory leads to an algebraic decay. The long-time behavior for the motion parallel and perpendicular to the wall is derived analytically and compared to numerical results.
 Then, we discuss the power spectral densities of the displacement and provide simple interpolation formulas. The theoretical predictions are finally  compared to recent experimental
observations.

\end{abstract}



\maketitle
\section{Introduction}

Understanding and controlling the motion of small colloidal particles suspended in fluids and confined to small volumes is essential in many applications in microfluidics~\cite{Squires:2005} and biophysics~\cite{Cicuta:2007}. Thermal fluctuations of the surrounding fluid agitate the colloids giving rise to Brownian motion.
Positioning the particle close to an interface or membrane and the  direct optical observation of  its trajectories
allows in principle to use the bead's motion as a sensor for the chemical and physical properties
of these two-dimensional surfaces. For example,  the presence of  surfactants~\cite{Blawzdziewicz:1999},  surface tension and elasticity~\cite{Felderhof:2006b,Bickel:2007}, or an adsorption layer~\cite{Felderhof:2006c} modifies the frequency-dependent mobility close to the interface.
To establish Brownian motion as a local reporter of frequency-dependent surface properties a detailed understanding of the  complex interplay of the hydrodynamic flow and the colloidal
thermal fluctuations is a prerequisite.
The simplest effect known as surface confinement, namely the anisotropic reduction of the diffusion coefficient close to a bounding  wall~\cite{Happel:LowReynolds,Lorentz:1907}, has been
directly observed experimentally only very recently~\cite{carbajal-tinoco:2007,Schaffer:2007}. To gain insight beyond the transport coefficients, one should record the time-dependent information, e.g. the mean-square displacement or the velocity autocorrelation, at times scales before the regime of simple diffusion is attained. Using weak optical trapping the
quasi-free Brownian particle's trajectory can be monitored interferometrically~\cite{Gittes:1998} with a time resolution of a few microseconds and a simultaneous positional sensitivity in the sub-nanometer regime~\cite{Lukic:2005,Lukic:2007}.

In the diffusive regime the momenta of the colloid are equilibrated with the surrounding fluid and are irrelevant for the Brownian motion. At very short times the particle's motion is ballistic as expected by Newton's laws. Hence, only after the colloid's initial momentum is transferred to the fluid the particle undergoes a random walk characterized by diffusion coefficients.
The time scale where this transition occurs  may be na\"{i}vely estimated by balancing the particle's inertia with the Stokes drag, $\tau_\text{p} = m_\text{p}/6 \pi \eta a$, where $m_\text{p}$ denotes the mass of the colloid, $a$ its radius,  and $\eta$ the shear viscosity of the fluid. Since momentum is conserved it can only be transported by vortex diffusion giving rise to another characteristic time scale
$\tau_\text{f} = a^2 \rho_\text{f}/\eta$, where $\rho_\text{f}$ is the density of the fluid. The time a vortex emitted from the colloid to reach the wall separated at a distance $h$ may be estimated by the characteristic time
$\tau_\text{w} = h^2 \rho_\text{f}/\eta$.  The dynamical information on how the diffusive regime is reached, is encoded in the velocity autocorrelation function (VACF) $C_\parallel(t) = \langle v_\parallel(t) v_\parallel(0)\rangle$ for the motion parallel to the surface and similarly perpendicular to the wall. In bulk the VACF
exhibits a power-law decay $t^{-3/2}$ at long times due to slow vortex diffusion.  This long-time tail is also expected  in the presence of
a wall for times up to $\tau_\text{w}$. For longer times this leading non-analytic behavior is canceled and a more rapid decay is expected, since the wall can carry away the momentum much faster.
An algebraic decay $t^{-5/2}$  for motion parallel  and a $t^{-7/2}$ long-time tail for motion perpendicular to
the wall has been predicted by Gotoh and Kaneda~\cite{Gotoh:1982} extending an earlier work of Wakiya~\cite{Wakiya:1964}. However, the result for the perpendicular motion is erroneous, as pointed out by Felderhof~\cite{Felderhof:2005}. He succeeded also in calculating  the  frequency-dependent mobility for all frequencies in both directions and thus giving an
analytic solution for the motion close to the wall up to Fourier transform. Later, he generalized this work to the case of a compressible fluid~\cite{Felderhof:2005b} and to a second bounding wall~\cite{Felderhof:2006}. Computer simulations for a colloidal particle confined between two walls demonstrate that the long-time behavior of the VACF is strongly affected
by the confinement~\cite{Hagen:1997,Pagonabarraga:1998,Pagonabarraga:1999,Frydel:2006,Frydel:2007}.

Recently, we have reported direct measurements of the velocity autocorrelation function  of a single colloid immersed in  water
close to a bounding wall~\cite{Jeney:2008}. We have observed that the time-dependent VACF becomes anisotropic and exhibits the non-algebraic tails that had been
anticipated much earlier~\cite{Wakiya:1964,Gotoh:1982,Felderhof:2005}.
To observe the particle for sufficiently long times close to a wall we used  a trap constraining the Brownian motion even more.
The optical trap introduces a harmonic restoring force $-K \vec{x}$, where $\vec{x}$ denotes the displacement from the trap center and $K$ the spring constant. Ignoring the wall a new time scale can be constructed $\tau_\text{k} = 6\pi \eta a/K$ characteristic of the positional equilibration in the trap.

Here we supplement our experimental results with a derivation of the theoretical description of the Brownian motion confined by a harmonic potential  and a  bounding wall. In particular, we discuss the complex interplay of the weak trap with hydrodynamic memory  originating from the obstructed vortex motion. We show
that even if the trapping time $\tau_\text{k}$ exceeds the characteristic time scales of particle momentum relaxation $\tau_\text{p}$, fluid momentum diffusion $\tau_\text{f}$, and
the wall-vortex reflection $\tau_\text{w}$, by several  orders of magnitude it still has a significant influence on the velocity autocorrelation function.
Section \ref{Sec:Felderhof} provides a brief introduction into the theoretical framework of Felderhof, which is extended by a harmonic restoring force in Sec.~\ref{Sec:trap}. An analytical  discussion of the emerging long-time tails is presented in Sec.~\ref{Sec:analytical} followed by a numerical study of the VACF and the power spectral density of the displacement in Sec.~\ref{Sec:numerical}. A comparison to our experimental results is shown in Sec.~\ref{Sec:experimental}.

\section{Felderhof's framework}\label{Sec:Felderhof}
In this Section we review the theoretical basis elaborated  by Felderhof~\cite{Felderhof:2005}
for the motion of a single colloid in the vicinity of a wall.
The fluid is treated as a continuum described by the Navier-Stokes equations, the colloid is modeled as an impenetrable  sphere
embedded in a viscous incompressible fluid. At the particle's surface the usual
no-slip boundary conditions are imposed. By the fluctuation-dissipation theorem
the velocity autocorrelation function is related to the frequency-dependent admittance~\cite{Zwanzig:1970}, hence it suffices to calculate the deterministic velocity response of the colloid to an external force.

The external force $\vec{R}(t)$  causes the motion of both the particle and surrounding fluid, which reacts by exerting a time-dependent drag force.
Newton's second law for the acceleration of the sphere reads after a temporal Fourier transform $-\text{i} \omega (m_{\text{p}}-m_{\text{f}}) \vec{U}_\omega = -\vec{F}_\omega+ \vec{R}_\omega$, where $\vec{U}_\omega$ is the velocity of the colloidal particle,  $m_{\text{p}}$ its  mass,  and $m_{\text{f}}$ denotes the mass of the displaced fluid.  The total \emph{induced  force} $\vec{F}_\omega$ corresponds to the frequency-dependent drag force on the particle up to the acceleration force $-\text{i} \omega m_{\text{f}} \vec{U}_\omega$ of a rigid sphere of
fluid of equal radius. We shall employ the \emph{point particle limit} where inhomogeneities of the flow on the scale of the colloid are ignored.

To calculate the total induced force one maps the problem in the presence of a bounding surface to a corresponding one in infinite space and employs
the generalized Fax\'{e}n theorem of Mazur and Bedeaux~\cite{Mazur:1974}: For an unbounded fluid the
frequency-dependent induced force $\vec{F}_\omega$ is
\begin{align}
\vec{F}_\omega &= \ekl \zeta(\omega) - \frac{3}{2}\text{i}\omega m_\text{f} \ekr (\vec{U}_\omega-\vec{v}'_\omega)  \, , \label{eq: mazur}
\end{align}
with $\zeta(\omega)= 6 \pi \eta a (1+ \sqrt{-\text{i} \omega \tau_\text{f}}) $
and $\vec{v}'_\omega$  denoting the unperturbed flow evaluated at the position of the particle $\vec{r}_0$. The branch cut is chosen such that $\sqrt{-\text{i} \omega \tau_\text{f}} = (1-\text{i}) \sqrt{\omega \tau_\text{f}/2}$.  Then consider as acting flow
\begin{equation}
\vec{v}'_\omega(\vec{r})= \vec{v}_\omega(\vec{r}) -\vec{v}_{0 \omega}(\vec{r}) \, ,
\end{equation}
where $\vec{v}_\omega(\vec{r}), \vec{v}_{0 \omega }(\vec{r})$ denote the solutions of the Navier Stokes equation in response to a point force
$\vec{R}_\omega$ acting at $\vec{r}_0$ in the presence of a bounding surface and in the infinite space.  Clearly,  $\vec{v}'_\omega(\vec{r})$ satisfies the homogenous Stokes equations.
The key idea is that  $\vec{v}_\omega(\vec{r})$ can be interpreted as a flow in infinite space resulting from the force $\vec{R}_\omega$ with $\vec{v}'_\omega(\vec{r})$ as an externally acting flow. Yet, for  flows in infinite space
 the generalized Fax\'{e}n theorem applies  and the induced force can be calculated easily.


By  linearity of the Navier-Stokes equation, the fluid response to a point force $\vec{R}_\omega$ acting at $\vec{r}_0$ is obtained by
\begin{equation}
\vec{v}_\omega(\vec{r})= G(\vec{r},\vec{r}_0) \cdot \vec{R}_\omega \, , \qquad
\vec{v}_{0 \omega}(\vec{r})= G_0(\vec{r}-\vec{r}_0) \cdot \vec{R}_\omega\, ,
\end{equation}
where $G(\vec{r},\vec{r}_0)$ and $G_0(\vec{r}-\vec{r}_0)$ are the corresponding (tensor) Green's functions.
Then the acting flow at the position of the particle reads $\vec{v}'_\omega=F(\vec{r}_0,\omega) \vec{R}_\omega$, where
\begin{equation}\label{eq:rft_def}
F(\vec{r}_0,\omega)= \lim_{\vec{r} \to \vec{r}_0} \kl G(\vec{r},\vec{r}_0) - G_0(\vec{r},\vec{r}_0) \kr \, ,
\end{equation}
defines the reaction field tensor. Its frequency dependence only enters via the ratio $h/\delta = \sqrt{\omega \tau_\text{f}/2}$, where $h$ is the distance to the wall and $\delta=\sqrt{2 \eta/\rho_\text{f}\omega }$ denotes the skin penetration depth.
Combining Newton's second law, Fax\'{e}n's theorem, and the reaction field approach one finds for the force balance
\begin{eqnarray}\label{eq:force_balance}
\lefteqn{- \text{i} \omega (m_\text{p}-m_\text{f}) \vec{U}_\omega =} \nonumber \\
&= & - \ekl \zeta(\omega) - \frac{3}{2}\text{i}\omega m_\text{f} \ekr  \ekl \vec{U}_\omega - F(\vec{r}_0,\omega) \vec{R}_\omega) \ekr+ \vec{R}_\omega \, .
\end{eqnarray}
Solving this equation for the particle velocity yields $\vec{U}_\omega= {\cal Y}(\omega) \vec{R}_\omega$, where the response function ${\cal Y}(\omega)$ is called the \emph{admittance tensor} and corresponds to  a frequency-dependent mobility.

Consider first the case where the wall is infinitely far away and the reaction field tensor vanishes. Then one finds for the admittance for infinite space
\begin{equation}\label{eq:Y0}
{\cal Y}_0(\omega) = \ekl
-\text{i} \omega m^* + \zeta(\omega) \ekr^{-1} \, .
\end{equation}
Here  $m^* = m_\text{p} + m_\text{f}/2$ can be interpreted as the \emph{effective mass} of the particle,
since  for a particle moving in an ideal fluid with constant velocity $\vec{U}$ the total kinetic energy including  the dragged fluid is given by $m^* \vec{U}^2/2$~\cite{Landau:1987}.

The general result in presence of a bounding wall can then be expressed as
\begin{eqnarray}
\lefteqn{ {\cal Y}(\vec{r}_0,\omega) = } \nonumber \\ &=&   {\cal Y}_0(\omega) \ekl 1 + 6 \pi \eta a \kl 1 + \sqrt{-\text{i}\, \omega \tau_\text{f}} + \frac{-\text{i}\, \omega \tau_\text{f}}{3} \kr F(\vec{r}_0,\omega) \ekr \, . \nonumber \\
\label{eq:Ywall}
\end{eqnarray}
Thus, all modifications due to the bounding wall are described by the reaction field tensor.
By symmetry only the motion parallel and perpendicular to the wall ${\cal Y}_{||}(\vec{r}_0,\omega)$, ${\cal Y}_{\perp}(\vec{r}_0,\omega)$ have non-zero components. Felderhof succeeded in providing a full analytical result for the frequency dependence of the corresponding reaction field tensors. We have verified  the sophisticated results for the parallel motion, Eq.~(3.9) in~\cite{Felderhof:2005} and the perpendicular motion~\cite{Felderhof:2006erratum}.
For further analysis, in particular, to evaluate the long-time tails in the velocity
autocorrelation function, it is sufficient to know their corresponding low-frequency expansions
\begin{equation}\label{eq:Fxxlow}
  F_{\parallel}(h, \omega)
 =  \frac{1}{6 \pi \eta h} \left( -\frac{9}{16} + v -\frac{9}{8} v^2 \nonumber + {\cal O}(v^3) \right)   \, ,
\end{equation}
and
\begin{equation}
F_{\perp}(h,\omega) = \frac{1}{6\pi \eta h} \left(-\frac{9}{8}+ v -
\frac{3}{8} v^2 +  {\cal O}(v^4)\right) \, . \label{eq:Fzzlow}
\end{equation}
where $v= (-\text{i} \omega \tau_\text{w})^{1/2}$.
Specializing to the stationary case, $\omega=0$, and inserting  into Eq.~(\ref{eq:Ywall}), one easily obtains
the zero-frequency admittance ${\cal Y}_{||,\perp}(h,0)$ recovering Lorentz's result~\cite{Lorentz:1907}
for the mobility close to a wall
\begin{equation}\label{eq:lorentz}
\mu_{\parallel} = \mu_0 \ekl 1 - \frac{9}{16} \frac{a}{h} \ekr \, , \, \,
\mu_{\perp} = \mu_0 \ekl 1 - \frac{9}{8} \frac{a}{h} \ekr \, ,
\end{equation}
where $\mu_0 = 1/6 \pi \eta a$ denotes the mobility in bulk.

\section{Influence of a harmonic trap}\label{Sec:trap}
In the framework presented above it is implicitly assumed that the distance $h$ between the particle and its bounding wall is time-independent. To realize this
situation experimentally and to prevent the particle from leaving  the detection region a trap is needed, which should be included to complete the model.
Already in bulk, the motion of a harmonically bound Brownian colloid differs drastically from a free particle, since at long enough times the restoring forces limit the particle's mean-square displacement to a finite value. The very first attempt by Uhlenbeck and Ornstein~\cite{Uhlenbeck:1930} to model the interplay of inertia of a particle, Stokes friction and harmonic restoring forces dates back to 1930 but ignores  hydrodynamic memory effects.
Only much later, Clercx and Schram~\cite{Clercx:1992} incorporated the fluid inertia generalizing the VACF for the free motion. In particular, they found that the long-time anomaly
changes from a $t^{-3/2}$ behavior for a free particle to a much more rapid decay according to  $t^{-7/2}$. Here, we combine the ideas of Felderhof for the free motion close to a confining wall with the externally acting harmonic restoring force.

Following the chain of arguments of Sec.~\ref{Sec:Felderhof} we include the
 harmonic restoring force in Newton's second law
\begin{equation}
- \text{i} \omega (m_\text{p}-m_\text{f}) \vec{U}_\omega = - \vec{F}_\omega - K \vec{x}_\omega + \vec{R}_\omega \,.
\end{equation}
Here $K$ denotes the spring constant of the optical trap, which we assume to be isotropic. The displacement of the particle relative to the trap center  at $\vec{r}_0$ will be eliminated in favor of the particle's velocity via
$\vec{U}_\omega =-
\text{i} \omega \vec{x}_\omega$. Since Fax\'{e}n's law and the reaction field tensor are unaffected by the presence of the trap, one can again eliminate the total induced force $\vec{F}_\omega$  and
merely supplement the force balance, Eq.~(\ref{eq:force_balance}), by the harmonic force
\begin{eqnarray}
\lefteqn{- \text{i} \omega (m_\text{p}-m_\text{f}) \vec{U}_\omega  -\frac{K}{\text{i} \omega} \vec{U}_\omega = }\nonumber \\
&= & - \ekl \zeta(\omega) - \frac{3}{2}\text{i}\,\omega m_\text{f} \ekr  \ekl \vec{U}_\omega - F(\vec{r}_0,\omega) \vec{R}_\omega) \ekr \nonumber +   \vec{R}_\omega \, .
\end{eqnarray}
In terms of the admittance tensor the previous relation may be written as
\begin{equation}
 \vec{U}_\omega = {\cal Y}(\vec{r}_0,\omega) \left[\frac{K}{\text{i}\, \omega} \vec{U}_\omega + \vec{R}_\omega \right]\, ,
\end{equation}
with the following interpretation: The particle's velocity  is still determined by the admittance tensor without trap, provided now the net force is considered as driving the system.
Solving for the velocity $\vec{U}_\omega = {\cal Y}^{(k)}(\vec{r}_0,\omega) \vec{R}_\omega$, one finds that the harmonic potential modifies the admittance
according to the simple rule
\begin{equation}\label{eq:Ytrap}
{\cal Y}^{(k)}(\vec{r}_0,\omega) = \left[ {\cal Y}^{-1}(\vec{r}_0,\omega) + K/-\text{i}\,\omega \right]^{-1} \, .
\end{equation}
This result includes Felderhof's result for zero trapping, $K=0$, as well as Clercx and Schram~\cite{Clercx:1992} if the wall is infinitely far away, ${\cal Y}(\vec{r}_0,\omega) \to
{\cal Y}_0(\omega)$. Again by symmetry, the matrix ${\cal Y}^{(k)}(\vec{r}_0,\omega)$ is diagonal with elements ${\cal Y}^{(k)}_{||}(\vec{r}_0,\omega)$
and ${\cal Y}^{(k)}_{\perp}(\vec{r}_0,\omega)$ for the force $\vec{R}_\omega$ parallel or perpendicular to the wall.

The trap introduces a new characteristic time scale $\tau_\text{k}$ into the problem: balancing the harmonic restoring force $K x$ with the
zero-frequency Stokes drag  $6\pi \eta a v$, one obtains $\tau_\text{k} = 6\pi \eta a/K = 1/\mu_0 K$. The reference drag force has been chosen as the one acting in bulk, although it is
clear that the presence of a boundary suppresses the hydrodynamic friction.
The physical interpretation of the dependences is that a stronger trap will lead to a faster relaxation to equilibrium resulting in a reduction  of $\tau_\text{k}$, whereas an increase of the Stokes drag induced by a larger viscosity leads to slowing down of the equilibration process.

In the derivation we have assumed that the trapping is isotropic, i.e. the spring constants are identical for all directions of the displacement. In case the trapping potential becomes anisotropic, the restoring force  is still given by $K \vec{x}_\omega$ where the spring constant $K$ has to be interpreted as a symmetric matrix. Then Eq.~(\ref{eq:Ytrap}) still holds, and if the principal axes of the trap includes the direction perpendicular to the wall, the matrix inversion is achieved by inverting the diagonal elements.

Let us compare  Eq.~(\ref{eq:Ytrap}) to the well known  high- and low-frequency limits. For high frequencies
the trap becomes increasingly irrelevant and the response is dominated by the inertia of the particle and the displaced fluid.
In the low-frequency regime the harmonic potential suppresses the admittance reflecting the fact that no static external force can induce a stationary motion of the particle in confinement.

\section{Analytical discussion of the Velocity autocorrelation function}\label{Sec:analytical}
 Here we focus on the velocity autocorrelation function (VACF) $C_\parallel(t) = \langle v_\parallel(t) v_\parallel(0) \rangle$ for the motion
parallel and perpendicular to the wall,  $C_\perp(t) = \langle v_\perp(t) v_\perp(0) \rangle$.  The VACF are connected to the admittances of the
previous Section via the fluctuation-dissipation theorem
\begin{equation}\label{eq:fluctuation-dissipation}
 {\cal Y}(\omega)  = \frac{1}{k_B T} \int_0^\infty\! \diff t \, \text{e}^{\text{i} \omega t} C(t) \, ,
\end{equation}
where $k_B T$ denotes the thermal energy.
This relation holds for all cases, i.e. parallel and perpendicular to the wall, with and without trap. To simplify notation we have suppressed
the dependence on the position $\vec{r}_0$.

Due to trapping the particle samples only a region close to the center of the trap and according to a stationary ensemble given
by the Gibbs-Boltzmann measure. In particular, time-dependent position correlation
functions like $\langle \vec{x}(t) \cdot \vec{x}(0) \rangle$ are well defined.
In the derivation for the admittances, the position of the particle was assumed to be fixed at $\vec{r}_0$. However, since the particle fluctuates and takes excursions from the center of
the trap of typical magnitude $\sqrt{k_BT/K }$, the spring constant  of the trap has to be strong enough in order to render these fluctuations negligible compared to the distance to the wall $h$.

The short-time evolution  of the velocity autocorrelation is inferred from the high-frequency behavior of the corresponding admittances. Obviously,
the harmonic restoring forces do not affect the initial value and they are still given by
\begin{align}
C_{\parallel}^{(k)}(t=0) &=   \frac{k_B T}{m^*} \left( 1 - \frac{a^3}{16 h^3} \right)\, , \nonumber \\
C_{\perp}^{(k)}(t=0)  &=   \frac{k_B T}{m^*} \left( 1 - \frac{a^3}{8 h^3} \right) \, .
\end{align}
Without a bounding surface  $C^{(k)}(t=0) = k_B T /m^*$~\cite{Clercx:1992} is recovered. The equipartition theorem
 suggests that the initial value should read $k_B T / m_{\text{p}}$ for all cases. However,
as discussed already by Zwanzig and Bixon~\cite{Zwanzig:1970}, there is an additional rapid decrease of the VACF if the finite compressibility of the fluid is taken into account.

\begin{table}
\begin{tabular}{l|l}
$6 \pi \eta a {\cal Y}(\omega) $ & leading nonanalytic term \\
\hline
0 & $ -(-\text{i} \omega \tau_\text{f})^{1/2}$ \\
$\perp$ &  $(-\text{i} \omega \tau_{\text{f}})^{3/2} \left[  - \frac{5}{9}+ \frac{\tau_{\text{p}}}{\tau_{\text{f}}} + \frac{1}{4} \sqrt{\frac{\tau_{\text{f}}}{\tau_{\text{w}}}} \left( 1 - \frac{9 \tau_{\text{p}}}{2 \tau_{\text{f}}} \right) \right]        $    \\
$\parallel$ & $(-\text{i} \omega \tau_{\text{f}})^{3/2} \left[ \frac{\tau_{\text{w}}}{\tau_{\text{f}}} - \frac{5}{9} + \frac{\tau_{\text{p}}}{\tau_{\text{f}}} + \frac{1}{8} \sqrt{\frac{\tau_{\text{f}}}{\tau_{\text{w}}}} \left( 1 - \frac{9 \tau_{\text{p}}}{2 \tau_{\text{f}}} \right) \right]        $    \\
$k$ &  $- (-\text{i} \omega \tau_{\text{f}})^{5/2} \tau_\text{k}^2/\tau_\text{f}^2 $                 \\
$k\perp$ &
$ - \frac{16}{9 (9-8 \sqrt{\tau_{\text{w}}/\tau_{\text{f}}})^2} (- \text{i} \omega \tau_\text{f})^{7/2} \frac{\tau_\text{k}^2}{\tau_\text{f}^2} \times $ \\
& $ \times
\left[  20 \frac{\tau_\text{w}}{\tau_\text{f}} \left( 1 - \frac{9\tau_\text{p}}{5\tau_\text{f}}\right) - 9 \sqrt{ \frac{\tau_\text{w}}{\tau_\text{f}}}
\left( 1 - \frac{9\tau_\text{p}}{2\tau_\text{f}}\right)
\right]
$         \\
$k\parallel$ & $ - \frac{32}{9 (9-16 \sqrt{\tau_{\text{w}}/\tau_{\text{f}}})^2} (- \text{i} \omega \tau_\text{f})^{7/2} \frac{\tau_\text{k}^2}{\tau_\text{f}^2} \times $ \\
& $ \times
\left[ -72 \frac{\tau_\text{w}^2}{\tau_\text{f}^2} + 40 \frac{\tau_\text{w}}{\tau_\text{f}} \left( 1 - \frac{9\tau_\text{p}}{5\tau_\text{f}}\right) - 9 \sqrt{ \frac{\tau_\text{w}}{\tau_\text{f}}}
\left( 1 - \frac{9\tau_\text{p}}{2\tau_\text{f}}\right)
\right]
$
\end{tabular}
\caption{Leading non-analytic term in the low-frequency expansion of the dimensionless admittances $6\pi \eta a {\cal Y}(\omega)$ for the free case (0), perpendicular ($\perp$)
and  parallel ($\parallel$) and again in the presence of the trap (k, k$\perp$, k$\parallel$)}\label{tab:admittances}
\end{table}
The low-frequency expansion of the admittance exhibits non-integer powers, which by tauberian theorems~\cite{Feller:Probability} correspond to  algebraic long-time decays in the VACF.
The leading long-time behavior of the normalized velocity autocorrelation, $C(t\to\infty)/C(0) \simeq    (\tau/t)^{\alpha}$, can be inferred from the
leading non-analytic contribution  in the admittance $Y(\omega \to 0) \simeq C(0) \tau (-\text{i} \omega \tau )^{\alpha-1} \Gamma(1-\alpha)/k_B T$. The physical origin of
the long-living correlations lies in the long-range fluid flow that is generated by momentum conservation in the Navier-Stokes equation. At low Reynolds number,
 transverse momentum can only be
transported away by vortex diffusion, hence there is a time-dependent growing length scale $R(t) \sim (\eta t/ \rho_\text{f} )^{1/2}$ characterizing how far momentum has penetrated into the fluid.
In the  simple case of unconfined free motion, the particle's initial momentum is then shared with a fluid volume $\sim R(t)^3$ leading to
a long-time behavior of $C(t)/C(0) \sim t^{-3/2}$.
For the free motion close to a bounding surface, the leading term
$ \sim t^{-3/2}$ is canceled and a more rapid
algebraic decay $t^{-5/2}$\cite{Felderhof:2005} is expected. Interestingly, the same algebraic decay occurs also in the disordered Lorentz gas~\cite{vanLeeuwen:1967,Ernst:1971b,Hoefling:2007}, a simple model for transport in porous media.
There the long-time memory arises since the particle remembers the presence of an obstructing wall for arbitrarily long time.
Confining the motion of a colloidal particle in bulk by 
a harmonic potential leads to a
more rapid decorrelation according to $t^{-7/2}$~\cite{Clercx:1992}.
 In the case of a trapped particle close to a wall,
a series expansion ${\cal Y}_\parallel^{(k)}(\omega), {\cal Y}_\perp^{(k)}(\omega)$ in powers of the frequency reveals a leading non-analytic
 term  $\omega^{7/2}$ resulting in a  $\sim t^{-9/2}$ tail.
To obtain this result it is  sufficient to know the reaction field tensors $F_\parallel(\vec{r}_0,\omega), F_\perp(\vec{r}_0,\omega)$
including  ${\cal O}(\omega)$ terms. Table \ref{tab:admittances} summarizes the leading non-analytic behavior of the admittances including all prefactors. The corresponding long-time tails in the normalized velocity autocorrelation function are displayed in Table \ref{tab:tails}, where for the walls only the leading term in $\tau_\text{w}/\tau_\text{f}$ is shown.

\begin{table}[tp]
\begin{tabular}{l|l}
$C(t)/C(0) $ & leading algebraic decay \\
\hline
0 & $  B \left(\frac{t}{ \tau_\text{f}} \right)^{-3/2}$ \\
$\perp$ & $ \frac{ 3}{2}   B \left( \frac{\tau_{\text{p}}}{ \tau_\text{f}} -\frac{5}{9} \right)\left(\frac{t}{ \tau_\text{f}} \right)^{-5/2}$\\
$\parallel$ & $\frac{3}{2} B  \left(\frac{\tau_{\text{w}}}{ \tau_\text{f}} \right)\left(\frac{t}{ \tau_\text{f}} \right)^{-5/2}$ \\
$k$ &   $  \frac{15}{4} B \left(\frac{\tau_\text{k}}{ \tau_\text{f}} \right)^2\left(\frac{t}{ \tau_\text{f}} \right)^{-7/2}$     \\
$k\perp$ &
 $\frac{105}{8} B \frac{\tau_\text{k}^2}{\tau_\text{f}^2} \left( \frac{\tau_\text{p}}{\tau_\text{f}} - \frac{5}{9} \right) \left(\frac{t}{ \tau_\text{f}} \right)^{-9/2}  $       \\
$k\parallel$ & $\frac{105}{8} B \frac{\tau_\text{w} \tau_\text{k}^2}{\tau_\text{f}^3} \left(\frac{t}{ \tau_\text{f}} \right)^{-9/2} $
\end{tabular}
\caption{The leading long-time behavior of the normalized velocity autocorrelation function $C(t\to \infty)/C(0)$ for the free case (0), perpendicular ($\perp$)
and  parallel ($\parallel$) and again in the presence of the trap (k, k$\perp$, k$\parallel$). For the cases of a bounding wall, only the leading contribution in $\tau_\text{w}/\tau_\text{f}$ is displayed.  The mass ratio is encoded in the constant $ B = (9 \tau_\text{p}/\tau_\text{f} +1)/18\sqrt{\pi}$. }
\label{tab:tails}
\end{table}
For the motion perpendicular to the wall, the true long-time behavior is masked by the next-to-leading term. This has been pointed out by Felderhof~\cite{Felderhof:2005} for the trap-free motion and here we
supplement the analysis also for the trapped dynamics.
In order to derive the next-to-leading term in the low-frequency expansion
the reaction field tensor $F_\perp(\omega,\vec{r}_0)$ has to be evaluated including the order $(-\text{i} \omega \tau_\text{f})^{5/2}$. Then the two leading nonanalytic terms in the far-field expansion
$\tau_\text{w}/\tau_\text{f} \gg 1$ read
\begin{eqnarray}
\lefteqn{ 6 \pi \eta a  {\cal Y}_\perp(\omega)  = } \nonumber \\
&= & (-\text{i} \omega \tau_\text{f})^{3/2}  \left[ \left( \frac{\tau_\text{p}}{\tau_\text{f}} - \frac{5}{9} \right)
-\frac{1}{10}\frac{\tau_\text{w}^2}{\tau_\text{f}^2} (-\text{i} \omega \tau_\text{f})  \right] \, ,
\end{eqnarray}
without harmonic restoring force, and
\begin{eqnarray}
\lefteqn{6 \pi \eta a  {\cal Y}_\perp^{(k)}(\omega) = } \nonumber \\
&=& (-\text{i} \omega \tau_\text{f})^{7/2} \frac{\tau_\text{k}^2}{\tau_\text{f}^2} \left[ \left( \frac{\tau_\text{p}}{\tau_\text{f}} - \frac{5}{9} \right)
-\frac{1}{10}\frac{\tau_\text{w}^2}{\tau_\text{f}^2} (-\text{i} \omega \tau_\text{f})  \right] \, ,
\end{eqnarray}
 including the trap. The second term becomes negligible with respect to the first one only at frequencies below $\omega \ll \tau_\text{f}/\tau_\text{w}^2$.
For the corresponding time correlation functions this implies a long-time behavior according to
\begin{eqnarray}\label{eq:Cperp_t}
\frac{C_\perp(t\to\infty)}{C_\perp(0)} &=& \frac{3}{2} B  \Big[  \left( \frac{\tau_\text{p}}{\tau_\text{f}}
- \frac{5}{9} \right) \left(\frac{t}{ \tau_\text{f}} \right)^{-5/2} \nonumber \\ 
& &+ \frac{\tau_\text{w}^2}{4 \tau_\text{f}^2} \left(\frac{t}{ \tau_\text{f}} \right)^{-7/2} \Big] \, ,
\end{eqnarray}

\begin{eqnarray}
\frac{C^{(k)}_\perp(t\to \infty)}{C^{(k)}_\perp(0)} &=& \frac{105}{8} B  \frac{\tau_\text{k}^2}{\tau_\text{f}^2}
\Big[ \left( \frac{\tau_\text{p}}{\tau_\text{f}} - \frac{5}{9} \right) \left(\frac{t}{ \tau_\text{f}} \right)^{-9/2} \nonumber \\
& &+
 \frac{9\tau_\text{w}^2}{20\tau_\text{f}^2}  \left(\frac{t}{ \tau_\text{f}} \right)^{-11/2}
\Big] \, .
\end{eqnarray}
The amplitude of the leading term depends on the ratio of the mass of the particle to the displaced mass of the fluid or equivalently on the ratio $\tau_\text{p}/\tau_\text{f} = 2 m_\text{p}/9 m_\text{f}$. For typical experimental conditions, e.g. a silica sphere in water, the amplitude is negative and one should expect the correlation function to approach zero from
below.
The presence of the factor $\tau_\text{w}/\tau_\text{f}$ renders the amplitude of the subleading term to a large quantity. Balancing both terms yields an estimate $\tau_\text{w}^2/\tau_\text{f}$  for the time one has to wait in order to observe the true long-time behavior. Hence the perpendicular motion without trap exhibits a zero crossing on
a time scale much longer than the na{\"\i}ve guess $\tau_\text{w}$. We shall see in the next Section that even a weak trap has strong implications for this intermediate time behavior.

\section{Numerical Results}\label{Sec:numerical}
\subsection{Velocity autocorrelation function}
The inverse Fourier transformation cannot be calculated analytically, since the  admittance tensors
are not simple elementary functions. A numerical Fourier transformation can easily be performed, and since the time-correlation functions are real
and even in time, a Fourier-Cosine transformation
is sufficient
\begin{eqnarray}
C(t) &=& \frac{2 k_B T}{\pi} \int_0^\infty \diff \omega \, \cos(\omega t) \text{Re }[ {\cal Y}(\omega)]\, .
\end{eqnarray}
Since the VACF exhibits long-time power-law behavior the admittance has to be sampled over many decades in frequency.
A conventional Fast Fourier Transform (FFT) is rather ill-suited to
achieve this and would also calculate $C(t)$ on an equidistant time grid  not convenient for our purposes. Hence, we apply a simple modified Filon algorithm~\cite{Tuck:1967}
with typically $10^3$ frequencies
per decade covering 30 decades to calculate $C(t)$ on a logarithmically equidistant grid containing 100 data points per decade. We employed Mathematica\textregistered\,
to evaluate the various special functions appearing in the reaction field tensors and spliced them together with a high-order series expansion
at low and high frequencies.
We have checked that
our numerical data reproduce the predicted tails over several orders of magnitude, with the exception of $C_\perp^{(k)}(t)$, where the next-to-leading tail dominates the figures.
\begin{figure}
\includegraphics[width=0.45\textwidth]{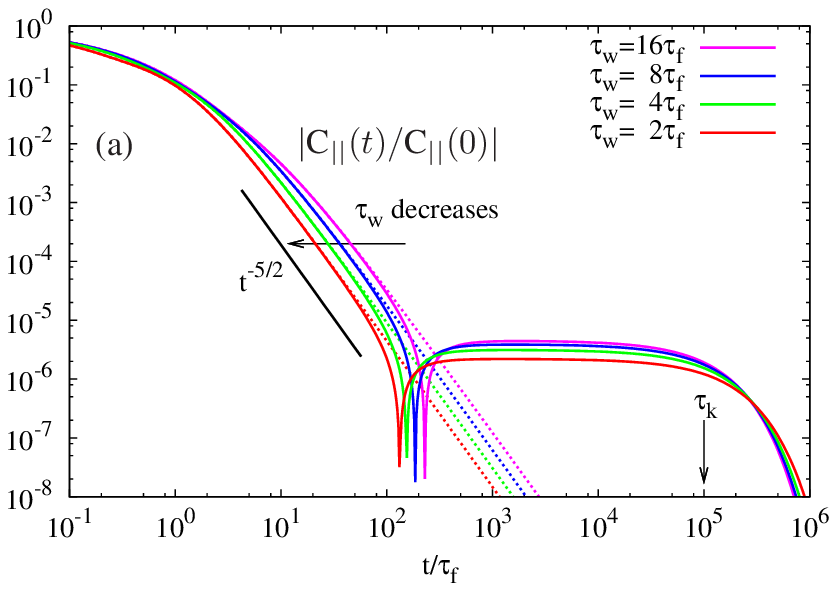}
\includegraphics[width=0.45\textwidth]{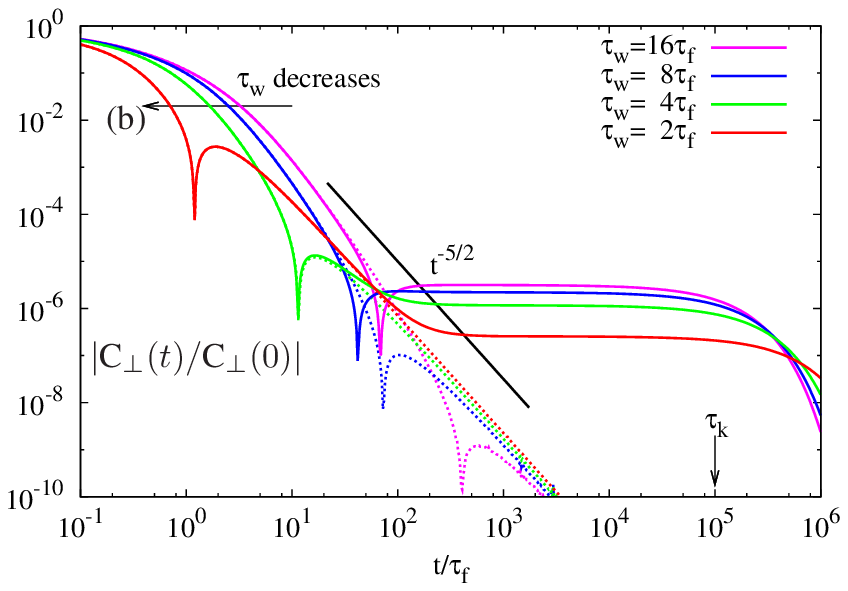}
\caption{(Color online) Double-logarithmic representation of the normalized
VACF for the motion parallel (a) and perpendicular (b) to the wall. The parameter $\tau_\text{p}/\tau_\text{f} = 0.5$ corresponds approximately to a silica bead
in water. Full lines correspond to weakly trapped particles $\tau_\text{f}/\tau_\text{k} = 10^{-5}$, whereas the dotted lines are without trapping ($\tau_\text{f}/\tau_\text{k} = 0$).
The wall is gradually approached, $\tau_\text{w}/\tau_\text{f}= 16,8,4,2$, corresponding to $h/a = 4, 2.83,2, 1.41$, and the initial decay shifts to the left. The straight lines are guides to the eye for the $t^{-5/2}$ power law.
 }
  \label{fig:Cwall}
\end{figure}

The velocity autocorrelation for the motion parallel to a wall is displayed in Fig.~\ref{fig:Cwall}a for a colloid gradually approaching the bounding surface.
Without trap the correlation functions remain positive for all times. The long-time behavior
exhibits the $t^{-5/2}$ decay with an amplitude that depends sensitively on the distance to the wall. For the weak  harmonic restoring force under consideration here, the VACF follows the
free behavior up to intermediate times; at longer times  the VACF is strongly influenced by the trap. In particular, the VACF exhibits two zeros as is the case also without the bounding wall. Note that the first zero is by orders of magnitude earlier than the characteristic time scale of the trap $\tau_k$, however much later than the Langevin momentum relaxation time $\tau_p$. Between the two zeros a negative flat plateau emerges
that extends to time scales longer than $\tau_\text{k}$. The very late decay is governed by the $t^{-9/2}$ tail of Table~\ref{tab:tails}, however its onset exceeds the range of the
figure.

The VACF in the direction perpendicular to the wall is exhibited in Fig.~\ref{fig:Cwall}b. For a typical experimental particle with momentum relaxation time $\tau_\text{p} = 0.5 \tau_\text{f}$ and without confining potential the curves exhibit a single zero. In this case the amplitude of the leading long-time behavior $t^{-5/2}$ becomes negative and the next-to-leading order $t^{-7/2}$ dominates at an intermediate
time interval, see Eq.~(\ref{eq:Cperp_t}). The correlation functions without trapping thus approach zero from the negative side by an algebraic decay that is fairly insensitive to the
distance from the wall. The zero on the other hand shifts rapidly to shorter times as the wall is approached. From Eq.~(\ref{eq:Cperp_t}) one infers an asymptotic scaling behavior $\sim\tau_\text{w}^2/\tau_\text{f}$ for the zero far away from the wall. For the two curves corresponding to the distant particle, $\tau_\text{w}/\tau_\text{f} = 8, 16$, the $t^{-7/2}$ decay is visible for intermediate times close to the zero, as has been reported earlier~\cite{Felderhof:2005}. For a weak trap $\tau_\text{f}/\tau_\text{k} =10^{-5}$ the curves follow the ones of the unconstrained motion down to a signal of $10^{-5}$ and start to  deviate strongly at later times. Again  a flat negative plateau characteristic of the weak restoring force is attained. The decay from this intermediate plateau occurs  at times large compared to $\tau_\text{k}$ followed by a rapid algebraic decay $t^{-9/2}$ without passing through another zero (not shown). For different parameters $\tau_\text{k}$, $\tau_\text{w}$ one finds also a scenario with three zero crossing.

To study the hydrodynamic memory effects for the colloidal motion close to the wall, we choose a trapping potential that is as weak as possible. Then the time scale $\tau_\text{k} = 6\pi \eta a/k$ is much larger than
the remaining time scales of the problem $\tau_\text{p}, \tau_\text{f}$, and $\tau_\text{w}$ and one is tempted to argue that the trap is irrelevant in the regime of interest. However, as is discussed
in Fig.~\ref{fig:Cwall} the trap has significant impact on the velocity correlation functions even if the characteristic time scales differ by orders of magnitude. The reason for such
a behavior is twofold: First, the hydrodynamic memory leads to a scale-free power-law long-time decay and  the parameters $\tau_\text{p}, \tau_\text{f}$, and $\tau_\text{w}$ determine merely the
amplitude of the algebraic behavior rather than a characteristic decay time. Second, for the point particle limit to apply accurately, $h/a \gtrsim 3$, and the VACFs become small at times
$t\gtrsim \tau_\text{w}$  where the tail $t^{-5/2}$ is expected to set in. Thus already a weak trap has a strong influence on the signal where the interesting hydrodynamic memory effect dominates the VACFs.

\subsection{Power spectral density}
An alternative way to investigate the interplay of fluid inertia  and a trap is to focus on  the power spectral density, an approach that has been pursued by Berg-S\o{}rensen \emph{et al.}~\cite{Berg-Sorensen:2004,Berg-Sorensen:2005}. There the fluctuating position $x(t)$ of the bead in a finite time interval $[-\text{T}/2,\text{T}/2]$ is decomposed into Fourier modes
$x_{\text{T}}(\omega) = \int_{-\text{T}/2}^{\text{T}/2}  x(t) \exp(\text{i} \omega t)\diff t$ where the angular frequencies $\omega$ are integer multiples of $2\pi/\text{T}$.
For long observation times, $\text{T}\to \infty$ the power spectrum $S(\omega)=\left\langle |x_\text{T}(\omega)|^2 \right\rangle/\text{T} $ becomes a quasi-continuous function for frequencies $\omega \gg 2\pi/T$. Since the corresponding velocities fulfill $v_\text{T}(\omega) = -\text{i} \omega x_\text{T}(\omega)$ up to irrelevant boundary terms, the power spectrum can be also expressed as $\omega^2 S(\omega)  = \left\langle |v_\text{T}(\omega)|^2 \right\rangle/\text{T} $. Then with the help of the Wiener-Khinchin theorem $ \omega^2 S(\omega) = \int   \langle v(t) v(0) \rangle \exp( \text{i} \omega t)\diff t$ and the fluctuation-dissipation theorem Eq.~(\ref{eq:fluctuation-dissipation}), the power spectral density  can be obtained from the admittance
\begin{equation}\label{eq:Srel}
S(\omega)= \frac{2 k_B T \, \mbox{\sf Re} \ekl {\cal Y}^{(k)}(\omega) \ekr}{\omega^2} \, .
\end{equation}
Since the trap modifies the admittance according to Eq.~(\ref{eq:Ytrap}), we find for the power spectrum
\begin{equation}
S(\omega)= \frac{2 k_B T\mbox{\sf Re} \ekl {\cal Y}(\omega)^{-1} \ekr}{ \kl \omega \, \mbox{\sf Re} \ekl {\cal Y}(\omega)^{-1} \ekr \kr^2+ \kl \omega \, \mbox{\sf Im} \ekl {\cal Y}(\omega)^{-1} \ekr + K \kr^2} \, .
\end{equation}
This result is again valid for all cases, i.e. in bulk, parallel and perpendicular to the bounding wall.
For the zero-frequency limit in the case of a limiting wall we find
\begin{equation}
S_{||}(0) = \frac{2k_B T}{\mu_{||} K^2} \, , \qquad
S_{\perp}(0) = \frac{2k_B T}{\mu_{\perp} K^2} \, ,  \label{eq:Stw}
\end{equation}
using the zero-frequency limit of the admittance at the wall. Without trap, the power  spectrum $S(\omega)$ would diverge for $\omega \to 0$
as can be seen from Eq.~(\ref{eq:Srel}), reflecting the fact that the particle can take excursions without bounds. The $1/K^2$ dependence on the trap strength  arises from
two aspects. First, the motion of the particle is harmonically confined leading to equilibrium fluctuations $\langle x^2 \rangle = k_B T/ K$. Second, the time scale to reach equilibrium
may be estimated by balancing the restoring force yielding $1/\mu K$. For the bulk motion this is identified with $\tau_k = 6\pi \eta a/K=1/\mu_0 K$, whereas close to the wall the reduction of the mobilities $\mu_\parallel, \mu_\perp$ due to the obstruction of the hydrodynamic flow has to be taken into account.

\begin{figure}[ht]
    \includegraphics[width=0.45\textwidth]{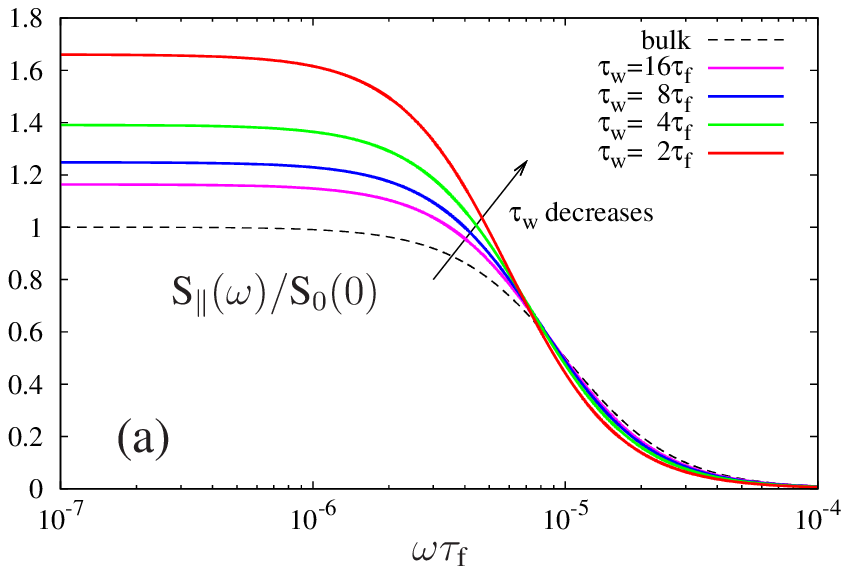}
    \includegraphics[width=0.45\textwidth]{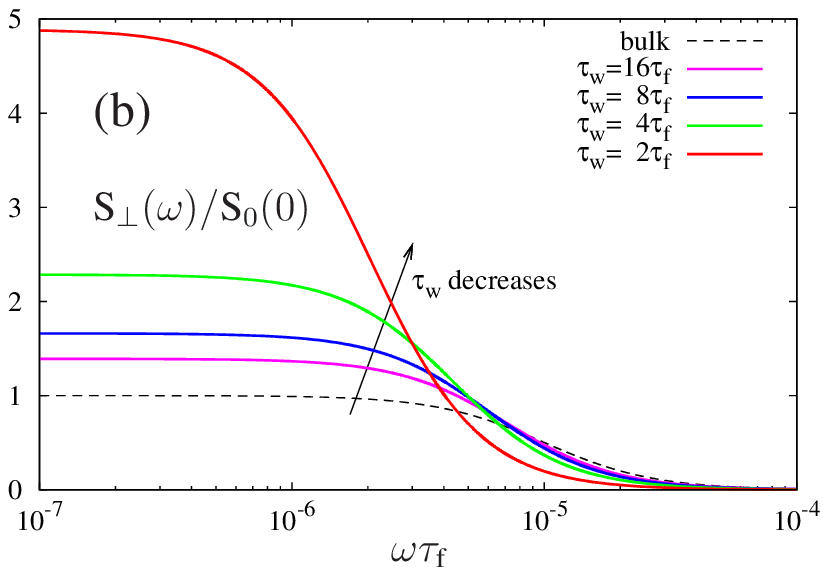}
\caption{(Color online) Semi-logarithmic representation of the power spectral density  for a weakly trapped Brownian particle $\tau_\text{f}/\tau_\text{k}= 10^{-5} $ close to a limiting wall for
 the ratio  $\tau_\text{p}/\tau_\text{f}=0.5$ ((a) lateral direction and (b) vertical direction).
The distances to the wall are the same  as in Fig.~\ref{fig:Cwall}.
The normalization is chosen as the zero-frequency limit of the power spectrum in bulk $S_0(\omega=0) = 2 k_B T/\mu_0 K^2$.
The increase of the initial value is due the wall according to Eq.~(\ref{eq:Stw}).
}
  \label{fig:Swall_tauk}
\end{figure}
At the scale of the trap relaxation rate $1/\tau_\text{k}$,  the main feature of the power spectral density as displayed in Fig~\ref{fig:Swall_tauk} is the saturation at a height given by Eq.~(\ref{eq:Stw}) at low frequencies. A rapid decrease of the power spectrum is manifest at higher frequencies. Ignoring inertial effects of the fluid and the particle, the power spectra assume a Lorentzian shape
\begin{align}\label{eq:lorentzian}
 S^{(L)}(\omega) = \frac{2 k_B T}{\mu K^2} \frac{1}{1+ (\omega/\mu K)^2} \, ,
\end{align}
where $\mu = \mu_0, \mu_\parallel, \mu_\perp$ is the corresponding mobility. For the weak traps employed here, this gives an accurate representation of the power spectral densities for
dimensionless frequencies $\omega\tau_\text{k}$  of order unity.

\begin{figure}[ht]
    \includegraphics[width=0.45\textwidth]{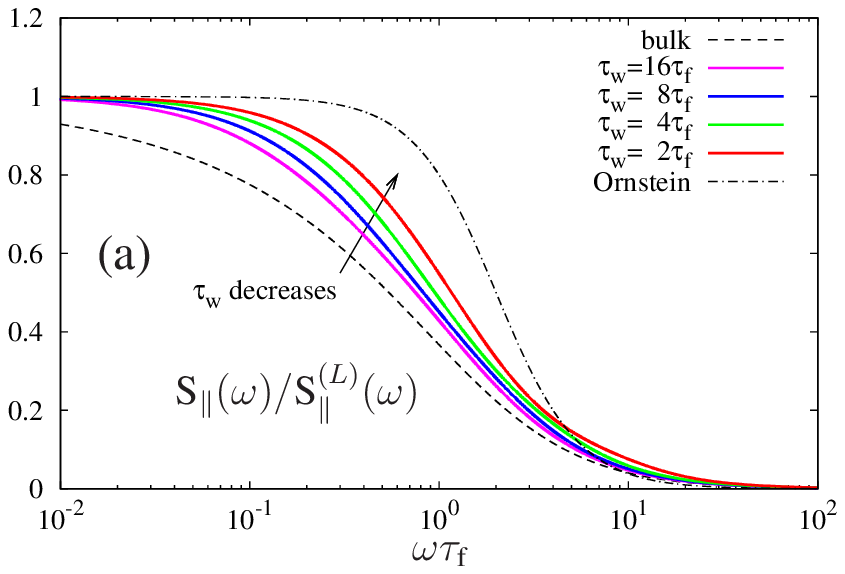}
    \includegraphics[width=0.45\textwidth]{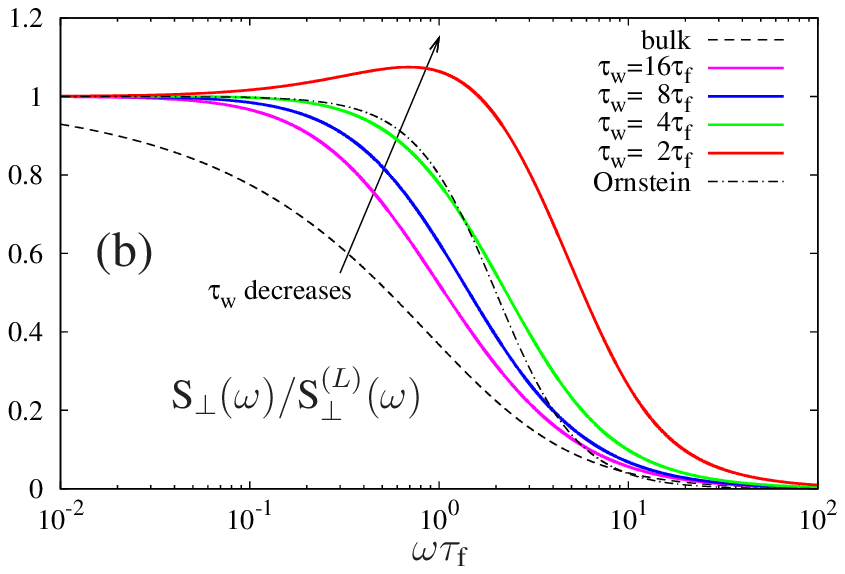}
\caption{(Color online) Power spectral density with respect to a  Lorentzian, Eq.~(\ref{eq:lorentzian}), in the frequency regime where fluid inertia plays a role for the parallel (a) and perpendicular motion (b). The parameters correspond to the ones of Fig.~\ref{fig:Swall_tauk}. Also included is the harmonic oscillator prediction according to Ornstein and Uhlenbeck.
}
  \label{fig:Swall_tauw}
\end{figure}

To highlight the deviations from the simple Lorentzian, we display the ratio $S(\omega)/S^{(L)}(\omega)$ for the parallel and perpendicular motion in Fig.~\ref{fig:Swall_tauw}. The
differences arise at the frequency scale $1/\tau_\text{f}$, where the motion of the hydrodynamic vortex sets in. At higher frequencies the full power spectrum decays more rapidly than a Lorentzian due the inertia of the particle and the fluid. Interestingly, the deviations are strongest in bulk and they fade out only slowly as the frequency is decreased. The wall suppresses these inertial effects consistent with the notion that the rigid interface carries away part of the particle's initial momentum. Hence the obstruction of the vortex pattern, manifest in the time-dependent VACF as reduction of a $t^{-3/2}$ to $t^{-5/2}$-tail, is also visible in the power spectral density by a faster approach to the Lorentzian shape for low frequencies $\omega\tau_\text{w} \lesssim 1$. Note that for the perpendicular motion the power spectral density becomes larger than a Lorentzian very close to the wall. This phenomenon is related to the change of sign of the long-time anomaly of the $t^{-5/2}$ in Eq.~(\ref{eq:Cperp_t}). The overshoot disappears if denser colloidal particles are used, i.e. $\tau_\text{p}/\tau_\text{f}$ is increased. However, the vanishing of the overshoot does not coincide with the sign change of the power-law tail at $\tau_\text{p}/\tau_\text{f} = 5/9$ but occurs at larger ratios of approximately 1.5 for the parameters used here. For comparison we have also included in both panels the harmonic oscillator prediction for the bulk motion following Ornstein-Uhlenbeck
\begin{equation}
 S^{(OU)}(\omega) = \frac{2 k_B T}{\mu_0 K^2} \frac{1}{(1-\omega^2 \tau_\text{p} \tau_\text{k})^2 + (\omega \tau_\text{k})^2} \, ,
\end{equation}
neglecting both, the added  mass due to the displaced fluid, as well as the vortex motion. As can be inferred from Fig.~\ref{fig:Swall_tauw}, the wall suppresses the hydrodynamic
memory, and for the parallel motion  the power spectral densities are typically in between  the bulk behavior where hydrodynamics is included and the Ornstein-Uhlenbeck shape.

For the weak trapping regime under consideration, the time scale $\tau_\text{k}$ is much larger than the other ones. Then one may assume that the processes of the relaxation of the momentum
to the surrounding fluid and the relaxation of the position in the trap are decoupled, which suggests to approximate the power spectral density as
\begin{equation}\label{eq:lorentz_approx}
S(\omega) \approx S^{(L)}(\omega) \frac{ \mbox{\sf Re} \ekl {\cal Y}(\omega) \ekr}{\mu} \, ,
\end{equation}
where again the appropriate mobilities $\mu = \mu_0, \mu_\parallel, \mu_\perp$ and similarly for the admittances ${\cal Y}(\omega)$ have to be inserted for the different cases.
We have checked that for the parameters used here this constitutes an excellent approximation for all frequencies. Consequently, Fig.~\ref{fig:Swall_tauw} essentially displays the
real part of the admittances for different distances to the wall. The reason that the curves superimpose is that the zero-frequency limit serves as a background that overlaps the interesting non-analytic behavior discussed in Sec.~\ref{Sec:analytical}. Consequently one can not use Eq.~(\ref{eq:lorentz_approx}) as input in the numerical Fourier transform since it does not result in the correct long-time behavior. Concluding, although the power spectral density allows for simple approximations on different frequency scales, their deviations encode subtle correlations that are manifest and more directly accessible in the time-dependent VACF.

\section{Comparison to Experiments}\label{Sec:experimental}

\begin{figure}[ht]
\includegraphics[width=0.45\textwidth]{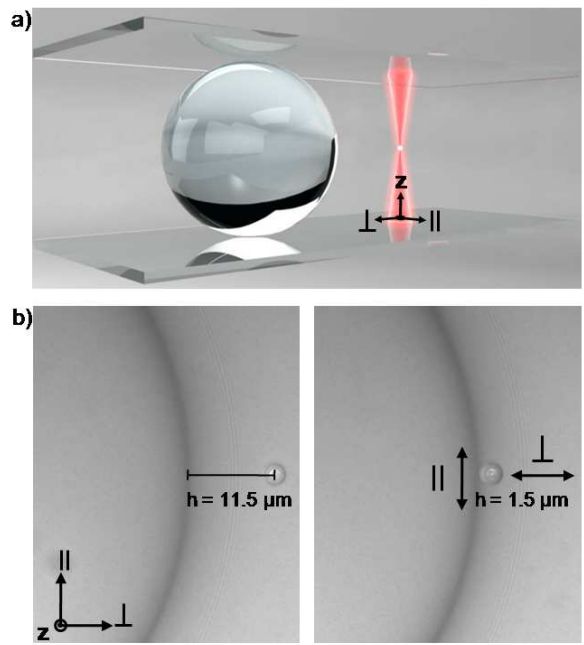}
\caption{\label{fig:scheme} (Color online)
(a) 3D lateral view of the experiment; spheres are drawn to scale. A silica particle of radius $a=$1.5\,\textmu{}m trapped by the laser focus is placed next to the surface of a
significantly larger silica sphere. This ~100\,\textmu{}m sphere is immobilized between
the two coverglass surfaces of the sample chamber.
(b) Optical image of the probing particle's position relative to the wall created by the big sphere.
The 3\,\textmu{}m probing particle was placed at a distance $h=$11.5\,\textmu{}m away
from the 100\,\textmu{}m sphere's surface and gradually approached. The velocity correlation functions, as well as the diffusion coefficients for the motion parallel and perpendicular to the wall are measured.
}
\end{figure}

Recently, we have performed  experimental tests for a small silica sphere  trapped by a laser focus in the vicinity of a surface~\cite{Jeney:2008}.
The trajectory of the particle is measured interferometrically~\cite{Gittes:1998} with a spatial resolution in the subnanometer range~\cite{Lukic:2005}.
We employ  an infrared ($\lambda=1064$\,nm) diode-pumped Nd:YAG laser (IRCL-500-1064-S, CrystaLaser, USA) with a cw output power of 500 mW.  The beam is first expanded  20$\times$ and then focused  using a 63$\times$ water-immersion objective
lens of a numerical aperture NA=$1.2$ resulting in a stable optical trap for the colloid.
 An InGaAs quadrant photodiode (G6849, Hamamatsu Photonics, Japan) is placed in the back focal plane of the condenser lens recording the modulation of the optical power
due to the displacement of the particle near the beam focus.
The photodiode signal is  amplified and digitized using a data acquisition card with a dynamic range of 12 bits. The detected positions
contain $N =10^7$ points separated by 2\textmu{}s, which corresponds to a sampling rate of $f_s = 500$\,kHz and a recording time of $t_s =20$\,s.

\begin{figure}[ht!]
\includegraphics[width=0.45\textwidth]{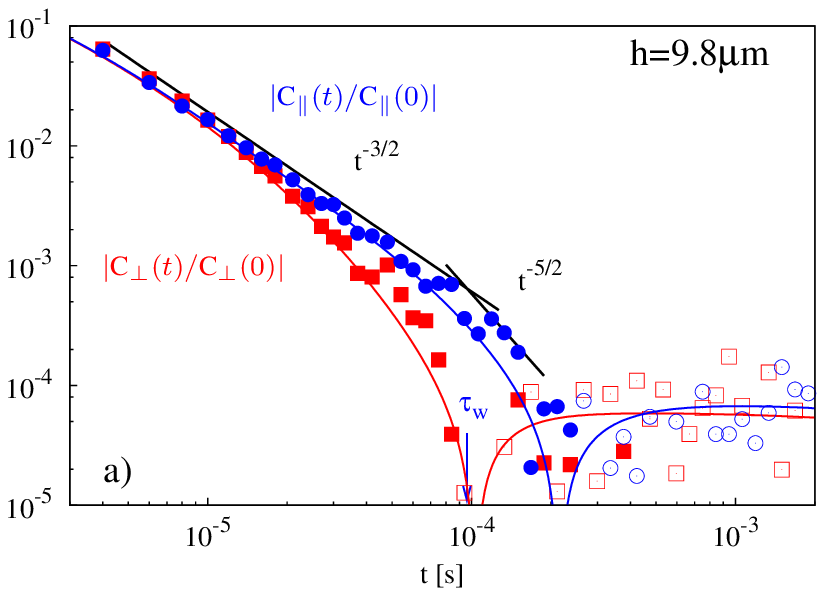}
\includegraphics[width=0.45\textwidth]{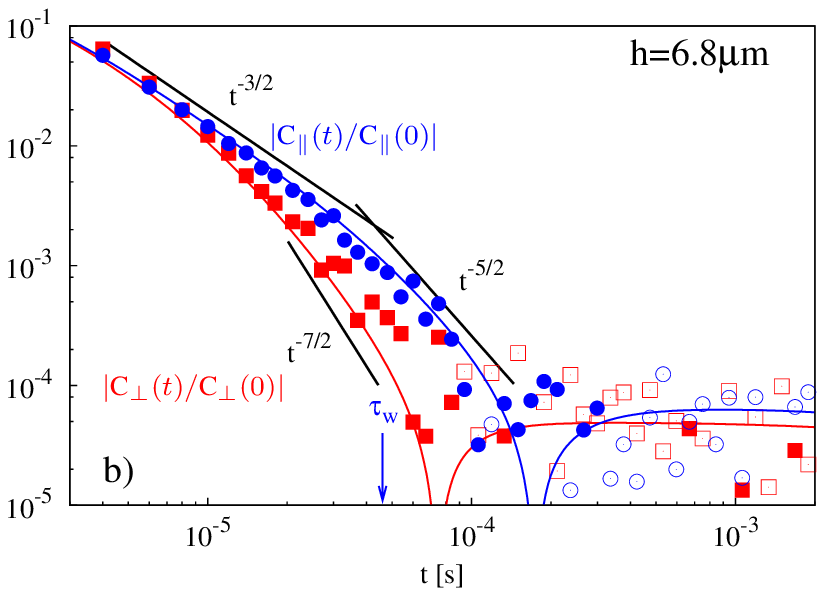}
\includegraphics[width=0.45\textwidth]{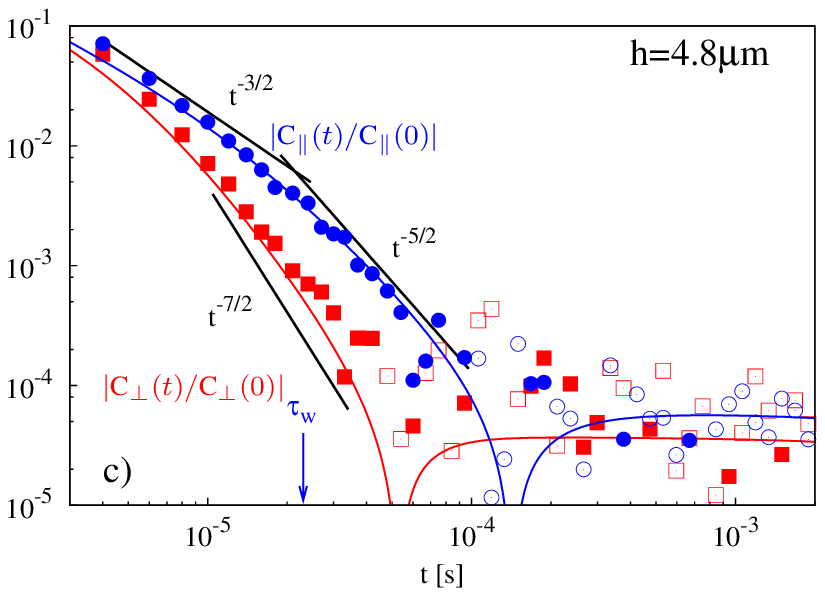}
\caption{\label{fig:LTT-wall} (Color online) Log-log plot of both normalized VACF, $C_\parallel(t)/C_\parallel(0)$ and $C_\perp(t)/C_\perp(0)$
for a sphere ($a=$1.5\,\textmu{}m, $\tau_\text{p} =1$\,\textmu{}s, $\tau_\text{f} = 2.25$\,\textmu{}s) trapped in a weak optical potential ($k$ $\approx$2\,\textmu{}N/m,
$\tau_\text{k}=14$\,ms). The increasingly anisotropic VACF is measured at three distances from the wall ($h$= 9.8, 6.8 and 4.8\,\textmu{}m, corresponding to $\tau_\text{w}$ =  96, 46, 23\,\textmu{}s, respectively).
Positive correlations are represented by full symbols, negative ones by open symbols.
The characteristic power-laws  are represented by thick lines as guide to the eye. }
\end{figure}

The Brownian particle is a silica sphere of radius $a=1.5$\,\textmu{}m and mass density $\rho_{\text{p}}=1.96$\,g/cm$^3$  immersed in water ($\rho_{\text{f}}=1$\,g/cm$^3$, $\eta=10^{-3}$\,Pa$\cdot$s)
at ambient temperature. The characteristic time scale corresponding to the particle's inertia $\tau_\text{p} = m_\text{p}/ 6 \pi  \eta a = 2 a^2\rho_\text{p} /9\eta = 1$\,\textmu{}s
is about half the one  of the fluid inertia $\tau_\text{f} = a^2 \rho_\text{f}/\eta =2.25$\,\textmu{}s. Using  optical tweezers, we approach the particle incrementally  towards
a sphere of diameter 100\,\textmu{}m, much larger than the size of our particle. Then sufficiently close to the large sphere the curvature can be ignored, and the particle undergoes Brownian motion close to a planar wall around the trap's center, see Fig.~\ref{fig:scheme}. The large sphere is immobilized since it is in close contact with the two
coverslides of our fluid chamber (size $\approx 2$\,cm $\times$0.5\,cm and thickness $\approx100$\,\textmu{}m). In this set-up we have equal sensitivity for the motion parallel and perpendicular to the surface under consideration, since both directions are perpendicular to the axis of the laser beam.
 The sample is mounted onto a piezo-stage, and the 100\,\textmu{}m sphere can be positioned at a
distance $h$ relative to the trapped particle by moving the piezo-stage in all three dimensions with a precision of $\approx1$\,nm.

To study the hydrodynamic memory effects induced by the presence of a wall
rather than the motion due to the trap confinement, we have  applied the weakest trapping force possible. Yet, the optical
trap should still be strong enough not to loose the particle from the laser focus during the experiment.
 Hence, we have  optimized the trapping strength in order to suppress the effects of the trap on the VACF  without loosing too much sensitivity at the detector. A series of experiments
revealed that $k$ $\approx$ 2\textmu{}N/m    fulfills all these requirements. Note that the corresponding time scale $\tau_k=14$\,~ms exceeds the
 parameters of the bulk motion $\tau_\text{p},
\tau_\text{f}$ by four orders of magnitude and the one of the wall $\tau_\text{w}$ by more than two orders of magnitude even for the farthest distance studied.
However, this weak trapping force has still significant influence on the VACF at the time scale where the hydrodynamic memory becomes apparent.

We have recorded the particle's position $r_\parallel(t_n), r_\perp(t_n)$ at equidistant instants of time $t_n := n\Delta t$, $\Delta t := 1/f_s$, $n\in \mathbb{N}_0$, and derive
coarse-grained velocities as $v_\parallel(t_n) := [r_\parallel(t_{n}+ \Delta t)- r_\parallel(t_n)]/\Delta t$. The velocity autocorrelation functions are then evaluated as time-moving averages $C_\parallel(t) =  N^{-1} \sum_{n=1}^N \Delta v_\parallel(t_n+t) \Delta v_\parallel(t_n)$, and
 similarly for the motion perpendicular to the wall. The normalized VACFs $C_\parallel(t)/C_\parallel(0)$ and $C_\perp(t)/C_\perp(0)$ are displayed in Fig.~\ref{fig:LTT-wall}
for different distances $h$ from the wall. For large separation $h=9.8$\,\textmu{}m both VACFs coincide within our error bars down to where the signal reaches the level of 1\%.
This regime is clearly dominated by bulk behavior and, in particular, the well-known algebraic decay $t^{-3/2}$ characteristic for the unconstrained vortex diffusion is recovered.
At the time scale $\tau_\text{w} = \rho_\text{f} h^2/\eta$ the vortex generated by the thermal fluctuations of the Brownian particle reaches the wall and
the correlation functions split as the motion becomes anisotropic. The wall leads to a  more rapid decay and for the parallel motion
a power law $t^{-5/2}$ enters the observation window . The signal of the perpendicular motion is anticorrelated at these time scales. Later times lead to only weak signals that we cannot resolve within our noise level of $10^{-4}$. For the VACFs the trap manifests itself in two zeros in the parallel motion, the second of which is outside of our observation window. The perpendicular motion exhibits a first zero which is shifted to earlier times by the presence of the optical trap --- a second zero induced by the harmonic restoring forces remains unobservable. Since for
silica in water  $\tau_\text{p}/\tau_\text{f} =0.44 < 5/9$,
the theory ignoring the trap also predicts an anticorrelated signal in the long-time behavior for the perpendicular motion. However, the negative signal we observe is dominated by the restoring force of the optical trap. Approaching the wall to $h=6.8$\,\textmu{}m and 4.8\,\textmu{}m, the characteristic time  $\tau_\text{w} =46$\,\textmu{}s, respectively 23\,\textmu{}s, decreases and the splitting into parallel and perpendicular motion shifts to earlier times. For the parameters chosen, the bulk behavior is dominated by the $t^{-3/2}$ tail,
and the splitting sets in at higher values of the correlation functions rendering it easier to observe. The crossover in  the parallel VACF from the bulk dominated behavior $t^{-3/2}$
to the wall dominated algebraic decay $t^{-5/2}$ becomes more and more pronounced. The zero due to the trap is dragged to shorter times since the surface constrains the colloid's motion
increasingly. The perpendicular component $C_\perp(t)$ decreases more  rapidly and practically enters our noise floor at the expected zero induced by the trap.
The asymptotic expansion for the long-time behavior of $C_\perp(t)$ suggests that a power law of $t^{-7/2}$ with positive amplitude should be present at intermediate times. Such a behavior
may  be inferred from the data for a narrow regime of times, but the presence of the trap suppresses  its amplitude by a factor of two.

\begin{figure}[ht]
\includegraphics[width=0.45\textwidth]{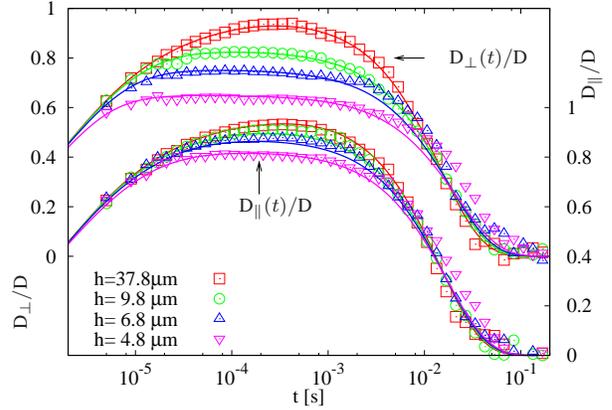}
\caption{\label{fig:Diffusion} (Color online) Normalized time-dependent diffusion coefficients, $D_\parallel(t)/D$ and $D_\perp(t)/D$, with the bulk diffusion constant $D=k_B T/6 \pi \eta a$ for the direction parallel and perpendicular to the wall at the same distances $h$ as in Fig.~\ref{fig:LTT-wall}.
The experimental data are represented by symbols and  the full lines correspond to the theoretical fits.
}
\end{figure}

Next, we define the time-dependent diffusion coefficients
\begin{eqnarray}
D_\parallel(t) = \int_0^t C_\parallel(t')\diff t'\, ,
\end{eqnarray}
and similarly for the motion perpendicular to the wall. From the recorded time series, $D_\parallel(t) = N^{-1} \sum_{n=1}^N [ r_\parallel(t_n+t)
-r_\parallel(t_n)]  v_\parallel(t_n)$ is directly evaluated. We have checked that this gives the same result as integrating the VACF. The diffusion coefficients $D_\parallel(t), D_\perp(t)$ corresponding to the distances
$h =9.8$, 6.8 and 4.8\,\textmu{}m of Fig.~\ref{fig:LTT-wall} as well as the motion in bulk ($h=37.8$\,\textmu{}m) are displayed in Fig.~\ref{fig:Diffusion}. The most prominent feature is a plateau extending to the time $\tau_\text{k}$ where the trapping becomes effective. The height of each plateau approaches the
diffusion coefficients parallel and perpendicular to the wall that are obtained by the Einstein-Smoluchowski relation from the mobilities, $D_\parallel = k_B T \mu_\parallel$, $D_\perp = k_B T \mu_\perp$. The reduction of the zero-frequency mobilities due to the wall in Eq.~(\ref{eq:lorentz}) ignores higher order correction terms in $a/h$ which can be calculated exactly~\cite{Happel:LowReynolds}. For the distances studied here $h/a>3$ the point particle limit is accurate within 2\% and the correction terms can be safely ignored. Since the VACF exhibits long-time tails, the approach to the plateau is slow. Ignoring the trap, we expect asymptotically for $t\to \infty$ in leading order in $a/h$
\begin{eqnarray}
 D_\parallel(t) = D \left( 1 - \frac{9a}{8h} \right) - D \frac{ \tau_\text{w}}{\tau_\text{f} \sqrt{4\pi}} \left( \frac{t}{\tau_\text{f}} \right)^{-3/2} \, ,
\end{eqnarray}
where $D= k_B T /6 \pi \eta a$ denotes the bulk diffusion constant. Even for the weak trapping used in the experiment the maximum in the time-dependent diffusion coefficient
deviates from the asymptotic value by a few percent. The maximum corresponds to the zero crossing in the VACF, which for the parallel motion is entirely due to the trap. At longer times  the diffusion coefficient $D_\parallel(t)$ decreases and eventually reaches zero.  The diffusion constant is given by the Green-Kubo relation $D_\parallel = \int_0^\infty C_\parallel(t') \diff t'$ or the long-time limit of the time-dependent diffusion coefficient.  Hence, at long times no diffusion occurs, which is  consistent with the observation that  the harmonic restoring forces localize the particle at sufficiently long times. Then the data at long times are sensitive to the precise value of the trap time $\tau_\text{k}$, and we have used this observation to optimize the fit for the trap stiffness. Since the laser power is held at a constant value for all distances, the trapping time should be identical for all data.
Figure ~\ref{fig:LTT-wall} shows that the terminal decay to zero can indeed be fitted by a single $\tau_\text{k}$, however for the closest distance $h=4.8$\,\textmu{}m the trap appears to be slightly weaker.
We attribute this observation to deformations of the laser field  at such close distances to the large sphere. Nevertheless, the overall agreement is excellent and validates the theoretical
approach on a quantitative level of a few percent. In particular, we conclude that even at the closest distance the point-particle limit is accurate not only for the stationary diffusion coefficient  but also for the time-dependent motion.

In the experimental set-up the planar wall has been realized by inserting a much larger spherical particle in the fluid chamber. Although  theoretical models have not addressed the Brownian motion close to a curved interface, some back-on-the envelope calculations may be made to confirm  the validity of the approximation. First, the zero-frequency mobility or steady diffusion constant will be modified and the suppression of mobility predicted by Lorentz, Eq.~(\ref{eq:lorentz}), should include also terms $h/R$, where $R$ denotes the radius of the large immobilized sphere. For the experiments
far away from the wall, $h=9.8$\textmu{}m, the suppression is less than 20\%, and $h/R \sim 1/10$ suggests that the curvature effect adds another 2\%. Close to the wall, $h =4.8$\textmu{}m, the suppression is already $35\%$, yet for the curvature an additional 1\% should be anticipated. One may also ask at what time scale $\tau_\text{curv}$ the effects of curvature should be manifest in the correlation functions. Assuming that curvature is relevant if the fictitious flat wall is separated from the immobilized sphere by more than $h$, simple geometric considerations lead to $\tau_\text{curv} = 2 \rho_\text{f} R h/\eta$. In our experiments these times are at least a factor of 20 larger than the corresponding $\tau_\text{w}$ and the signal is indistinguishable from noise.
\section{Conclusion}

Optical trapping interferometry  allows  monitoring the motion of a single colloidal particle on time scales where  its momentum plays an important role. Due to the hydrodynamic memory of the fluid friction, the velocity autocorrelation function (VACF) exhibits an algebraic long-time decay rather than an exponential relaxation. If the
colloid is placed in close vicinity to a bounding wall the vortex diffusion is significantly hindered leading to a more rapid decay of the VACF. Although weak optical trapping is employed,
where the trap relaxation time scale exceeds the ones of the fluid by orders of
magnitude, the influence of the confining harmonic potential becomes significant.
At times where the algebraic decay due to the wall should be visible the trap introduces additional features in the signal which have to be disentangled carefully.
To analyze experimental data the full frequency dependence of the admittance including the trap has to be used to obtain a consistent interpretation.

Our study should be useful to analyze experiments of colloidal particles close to an interface and/or in visco-elastic fluids in a straightforward way. Employing  the  time-domain rather than the frequency-domain should make the 
  interpretation of data on visco-elastic solutions, e.g. conducted by   Atakhorrami \emph{et al} \cite{Atakhorrami:2006}, more transparent and significant, and improve  the trapping force calibration method
  based on the power spectral density suggested by Fischer and Berg-S\o{}rensen~\cite{Fischer:2007}.
In visco-elastic media the viscosity becomes itself
frequency-dependent which is readily incorporated in the admittances. The velocity autocorrelation functions can be calculated numerically once the frequency dependence is known.
Provided the experimental data exhibit little noise one may determine the frequency-dependent elastic moduli by adjusting the numerically generated curves to the experiment. Similarly,
the admittances change as the properties of the interface change by adding surfactants, by capillary fluctuations, surface viscosity, etc. If the admittances for each case are known one may use the fluctuating bead as a probe for the local environment and determine material properties on scales ranging from nano- to  micrometers.


\acknowledgments
We thank F. H{\"o}fling for discussions and help to implement the Filon algorithm.
SJ thanks Ecole
Polytechnique F\'ed\'erale de Lausanne (EPFL) for funding the
experimental equipment. This work is supported by the Swiss National Foundation under grant no. 200021-113529.
 TF gratefully acknowledges support by the Nanosystems Initiative Munich (NIM).


\end{document}